\documentclass[amssymb,amsmath,aps,groupedaddress,reprint,superscriptaddress,twocolumn,showpacs]{revtex4}

\usepackage{graphicx}% Include figure files
\usepackage{dcolumn}% Align table columns on decimal point
\usepackage{bm}% bold math
\usepackage{amsmath}

\begin{document}

\title{Generation of entangled photons in graphene in a strong magnetic field}
\author{Mikhail Tokman}
\affiliation{Institute of Applied Physics, Russian Academy of Sciences, Nizhny Novgorod, Russia}
\author{Xianghan Yao}
\affiliation{Department of Physics and Astronomy, Texas A\&M
University, College Station, TX 77843 USA}
 \author{Alexey Belyanin}
 \email{belyanin@tamu.edu}
\affiliation{Department of Physics and Astronomy, Texas A\&M
University, College Station, TX 77843 USA}

\date{23 September 2012}

\begin{abstract}

Entangled photon states attract tremendous interest as the most vivid manifestation of nonlocality of quantum mechanics and also for emerging
applications in quantum information. Here we propose a mechanism of generation of polarization-entangled photons, which is based on the nonlinear optical interaction (four-wave mixing) in graphene placed in a magnetic field.  Unique properties of quantized electron states in a magnetized graphene and optical selection rules near the Dirac point give rise to a giant optical nonlinearity and a high rate of photon production in the mid/far-infrared range. 
\end{abstract}


\maketitle

To date, the most widely used method of
generating entangled photons is based on the spontaneous parametric down-conversion in a nonlinear crystal possessing a second-order
nonlinearity \cite{kwiat,nature2012}. In this process, a photon from a strong pump field at frequency $\omega_p$ splits into two signal photons, $\omega_p = \omega_1
+ \omega_2$ which can be entangled in polarization, frequency, and wave vector. Entanglement in the polarization degree of freedom is the most
convenient one for applications. Another way to generate quantum-correlated photons through a parametric nonlinear optical process
is spontaneous four-wave mixing in the optical fibers, in which two pump photons are converted into two signal photons, $2\omega_p = \omega_1 +
\omega_2$, utilizing a third-order nonlinearity of silica \cite{fan-opn07}. This process is obviously compatible with fiber communication
technologies, although it does not directly lead to polarization entanglement. In order to achieve the latter, one needs to use two pumps with crossed
polarizations and apply additional signal processing.  In both nonlinear processes the photon pair production efficiency is very low. An alternative approach utilizing the radiative decay of biexcitons in semiconductor quantum dots   \cite{stevenson,mohan,dousse} allows photon pairs to be generated on demand but requires cooling down to liquid helium temperatures.

Graphene has unusual electronic and optical properties stemming from linear, massless dispersion of electrons near the Dirac point and the chiral character of electron states \cite{castroneto,nair}. Magnetooptical properties of graphene and thin graphite layers are particularly peculiar, showing multiple absorption peaks and unique selection rules for transitions between Landau levels \cite{sadowski,abergel2007,prl,kono2012}. Recent progress in growing high-quality epitaxial graphene and graphite with high room-temperature mobility and strong magnetooptical response attracted a lot of interest and paved the way to new applications in the infrared optics and photonics \cite{orlita2008,orlita2009,crassee}.  The time is ripe to explore the nonlinear and quantum optical properties of a magnetized graphene and their applications. We have recently shown that graphene placed in a magnetic field possesses perhaps the highest infrared  optical nonlinearity among known materials \cite{prl}. Here we argue that an extremely strong nonlinearity of graphene in combination with its peculiar properties of the Landau levels open new avenues for generation of the nonclassical light states, in particular polarization-entangled photons. A similar mechanism of  photon entanglement may exist in topological insulators where the surface states have a Dirac-cone dispersion and demonstrate similar properties of magneto-optical absorption. 
%%%%%%%%%%%%%%%%%%%%%%%%%%%%%%%%%%%%  figure 1 %%%%%%%%%%%%%%%%%%%%%%%%%%%%%
\begin{figure}[htb]
\centerline{
\includegraphics[width=7cm]{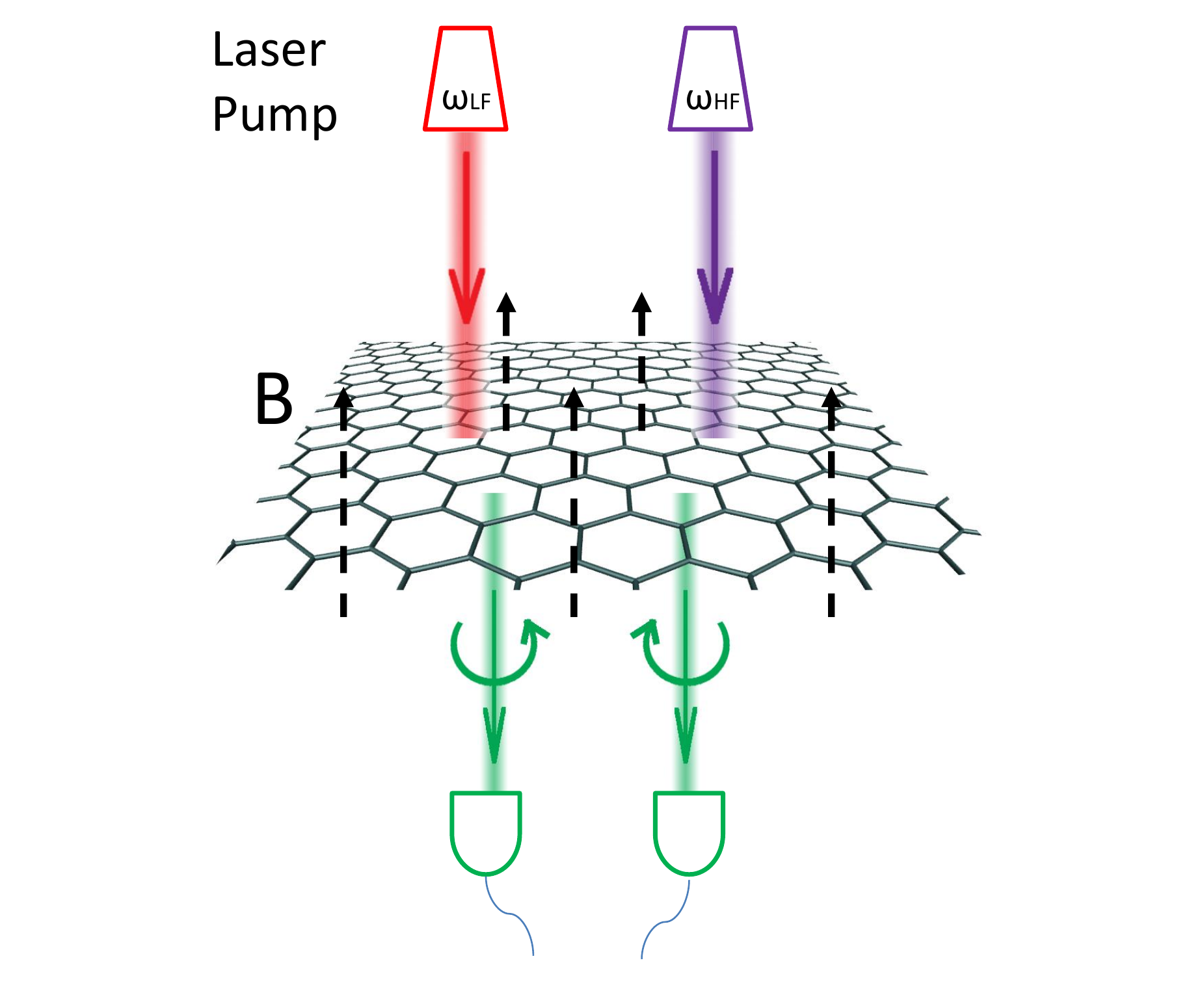}}
\caption{ Geometry of the proposed experiment. Two pump fields at frequencies $\omega_{HF}$ and $\omega_{LF}$ normally incident on a sheet of graphene placed in a magnetic field B generate entangled photons with opposite sense of the circular polarization. }
\end{figure}
%%%%%%%%%%%%%%%%%%%%%%%%%%%%%%%%%%%%%%%%%%%%%%%%%%%%%%%%%%%%%%%%%%%%%%%%%%%%
%%%%%%%%%%%%%%%%%%%%%%%%%%%%%%%%%%%%  figure 2 %%%%%%%%%%%%%%%%%%%%%%%%%%%%%
\begin{figure}[htb]
\centerline{
\includegraphics[width=10cm]{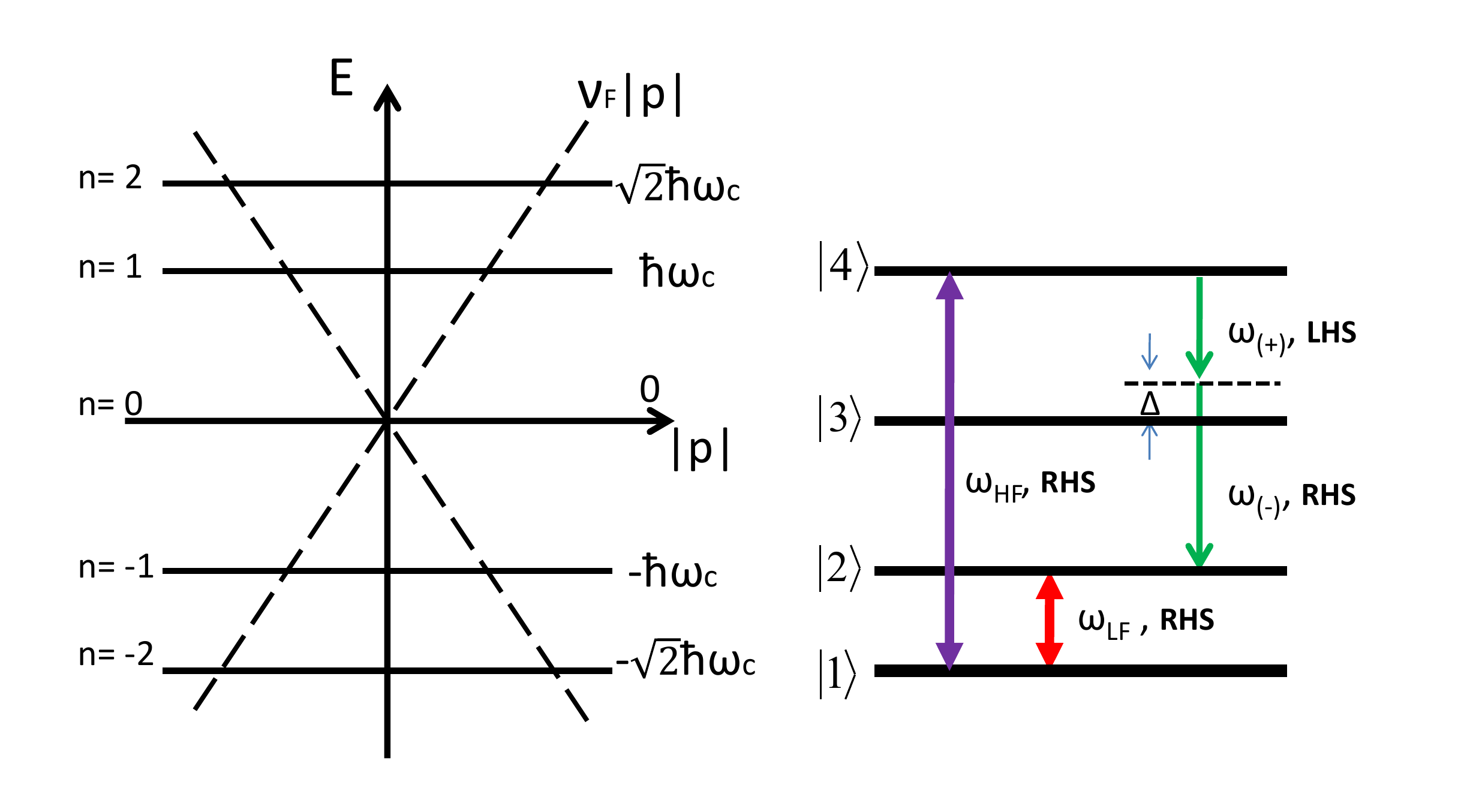}}
\caption{ Energy levels and optical transitions involved in resonant parametric generation of entangled photons in graphene. Left: Landau levels near the Dirac point superimposed on the linear electron dispersion without the magnetic field. Right: A scheme of the entangled photon
generation process in the four-level system of LLs with energy quantum numbers n = -2, -1, 0, 1 that were renamed as states 1,2,3, and 4 for
convenience of notation. Right-hand side (RHS) or left-hand side (LHS) circular polarization of light indicated on the figure corresponds to the allowed transitions.}
\end{figure}
%%%%%%%%%%%%%%%%%%%%%%%%%%%%%%%%%%%%%%%%%%%%%%%%%%%%%%%%%%%%%%%%%%%%%%%%%%%%

The proposed scheme is shown in Figs.~1 and 2. Here the energies of the Landau levels for electrons near the Dirac point are given by $\varepsilon_n={\rm
sgn}(n)\hbar\omega_c \sqrt{|n|}$ , where $ n = 0, \pm 1, \pm 2 ...$, $\omega_c = \sqrt{2}\upsilon_F/l_c$, $\upsilon_F \approx 10^8$ cm/s the electron
Fermi velocity, and $l_c = \sqrt{\hbar c /eB}$ the magnetic length. We assume that the graphene is biased or doped so that the Fermi level is between
the states with n = -2 and n = -1, i.e. the state n = -2 is occupied and the states above are empty in the absence of pumping. Two incident strong
pump fields at frequencies $\omega_{HF}$ and $\omega_{LF}$ resonant to the transitions from n = -2 to n = 1 and from n = -2 to n = -1 respectively, generate
two signal fields with opposite sense of the circular polarization at frequencies $\omega_{(-)}$ and $\omega_{(+)}$ that are close to resonance with transitions
from n = -1 to 0 and from n = 0 to 1. Note that these transitions have the same energy. Therefore, the presence of the unshifted $n=0$ Landau level enables convenient entanglement in the polarization degree of freedom for two photons with nearly equal energies. All transition frequencies are easily tunable with a magnetic field.

The polarizations for the allowed transitions are indicated in Fig.~2. Here LHS and RHS denote left-hand and right-hand circularly polarized light
with polarization vectors in the (x,y) plane of the graphene defined as ${\bf e_{(\mp)}}= \left( {\bf x_0} \mp i \bf {y_0} \right)/\sqrt{2}$, respectively. Peculiar
selection rules for graphene, $\Delta |n| = \pm 1$  as opposed to $\Delta n = \pm 1$ for electrons with usual parabolic dispersion, allow transitions
with a large change in the principal quantum number n, such as the transition from n = -2 to 1. The dipole matrix elements of the allowed transitions
$d_{mn} \sim \hbar e \upsilon_F / (\varepsilon_n - \varepsilon_m)$ grow fast ($\sim\lambda$) with increasing wavelength, and reach a large magnitude
in the mid/far-infrared range; e.g. $|d|/e = 13$ nm for the $n = 0$ to $n = 1$ transition in the field of 1 Tesla (at wavelength of 34 $\mu$m). This enables an extremely high resonant third-order
nonlinearity \cite{prl}. Note also that the states $n$ = -1, 0, and 1 have low population when the intensities of the optical pumps are below saturation and $\hbar \omega_c \gg k_B T$.
These factors lead to a high rate of photon generation and high signal to noise ratio of entangled photons.

In order to determine the optimal conditions for entanglement and the photon generation rate, we solve coupled equations for Heisenberg
operators of the electron and signal photon fields, assuming that the strong pump fields are classical. 
Consider quasiparticles ("electrons") on Landau levels described by stationary states $|m\rangle$ and energy levels $\varepsilon_m$. After introducing creation and annihilation operators of an electron, $\hat{a}^\dag_m|0\rangle = |m\rangle, \hat{a}_n|n\rangle = |0\rangle$, one can define a coordinate-dependent density operator,
\begin{equation}
\hat{\rho}_{mn}(\textbf{r},t) = \frac{1}{\Delta V_r}\sum_{j}\hat{a}^\dag_{j;n}(t)\hat{a}_{j;m}(t)
\label{rhodef}
\end{equation}
where the index $j$ numerates individual electrons and the summation is carried over all electrons within a small volume $\Delta V_r$ in the vicinity of a point with radius-vector $\textbf{r}$. The density operator in Eq.~(\ref{rhodef}) is normalized to the total electron density $N(\textbf{r})$ according to $\langle \Psi_E | \sum_m \hat{\rho}_{mm} | \Psi_E \rangle = N(\textbf{r})$ where $|\Psi_E\rangle = \sum_m C_m |m\rangle$ is the wave function of the electron subsystem satisfying $\sum_m |C_m|^2 = 1$. Assuming that the operators in different points of space commute with each other, the commutation relations become
\begin{equation}
\lbrack \hat{\rho}_{qp}(\textbf{r}), \hat{\rho}_{mn}(\textbf{r}^\prime)\rbrack = \delta(\textbf{r}-\textbf{r}^\prime)(\hat{\rho}_{mp}(\textbf{r})\delta_{qn} - \delta_{mp}\hat{\rho}_{qn}(\textbf{r})). 
\label{commut} 
\end{equation}
 Using the density matrix defined in Eq.~(\ref{rhodef}), one can write the Heisenberg operator of any physical quantity $x(\textbf{r},t)$ as
$ \hat{x} = x_{nm}\hat{\rho}_{mn}(\textbf{r},t)$.
In particular, the optical polarization is given by $\hat{\textbf{P}}(\textbf{r},t) = \textbf{d}_{nm}\hat{\rho}_{mn}$ where $\textbf{d}_{mn}$ is the dipole matrix element.

The Heisenberg-Langevin equation for the density operator takes the form
\begin{equation}
\dot{\hat{\rho}}_{mn} = -\frac{i}{\hbar}\left(\hat{h}_{mv}\hat{\rho}_{vn} - \hat{\rho}_{mv}\hat{h}_{vn}\right) + \hat{R}_{mn}(\hat{\rho}_{mn}) +
\hat{F}_{mn},
\label{rho}
\end{equation}
independently on whether $\hat{a},\hat{a}^\dag$ operators obey the commutation relations for fermions or bosons.
In Eq.~(\ref{rho}) $\hat{h}_{nm} = \varepsilon_n\delta_{nm} - \textbf{d}_{nm}\hat{\textbf{E}}(\textbf{r},t)$ is the matrix element of the Hamiltonian operator $\hat{H} =
\hat{h}_{nm}\hat{a}^{+}_n\hat{a}_m$ describing interaction with the electric field $\hat{\bf{E}}(\textbf{r},t)$ in the dipole approximation and $\hat{R}_{mn}$ the relaxation operator, for which we will choose the simplest form $\hat{R}_{m\ne n} = -\gamma_{mn}\rho_{mn}$. 
 
 The Langevin noise operator $\hat{F}_{mn}$ satisfies $\hat{F}_{mn} = \hat{F}^\dag_{nm}$ and $\langle \hat{F}_{mn} \rangle = 0$. Here the averaging
$\langle ... \rangle$ is taken both over the reservoir and over the initial state $|\Psi_E\rangle$ of the electron system. Its commutator is equal to
\begin{equation*}
\lbrack\hat{F}_{mn}(\textbf{r},t), \hat{F}^\dag_{mn}(\textbf{r}^\prime, t^\prime)\rbrack = 2\gamma_{mn}(\hat{\rho}_{nn} -
\hat{\rho}_{mm})\delta(t-t^\prime)\delta(\textbf{r}-\textbf{r}^\prime)
\end{equation*}
and correlation relations for its spectral components are 
\begin{equation*}
\begin{array}{cc}
\langle \hat{F}^\dag_{\omega;mn}(\textbf{r}^\prime) \hat{F}_{\omega^\prime;mn}(\textbf{r})\rangle & = \displaystyle  \frac{\gamma_{mn}}{\pi}\langle \hat{\rho}_{mm} \rangle \delta(\omega -
\omega^\prime)\delta(\textbf{r} - \textbf{r}^\prime), \\
\langle \hat{F}_{\omega^\prime;mn}(\textbf{r}) \hat{F}^\dag_{\omega;mn}(\textbf{r}^\prime)\rangle &= \displaystyle   \frac{\gamma_{mn}}{\pi}\langle \hat{\rho}_{nn} \rangle \delta(\omega -
\omega^\prime)\delta(\textbf{r} - \textbf{r}^\prime). 
\end{array}
\end{equation*}
 Their derivation follows standard steps described in \cite{landau} for any stationary random delta-correlated process, but without making the assumption of a thermal equilibrium. In the limiting case of a thermal equilibrium, $\langle\hat{\rho}_{nn}\rangle / \langle\hat{\rho}_{mm}\rangle = \exp{(\hbar\omega_{nm}/k_BT)}$, one can derive from the above correlation relations that the fluctuation component of the polarization generated by the Langevin noise satisfies the fluctuation-dissipation theorem \cite{landau,rytov}.  

For a monochromatic electric field of a given field mode propagating in a dispersive medium with refraction index $n(\omega)$, $\hat{\textbf{E}} =
\hat{\textbf{E}}_0e^{-i\omega t + ikz} + \hat{\textbf{E}}^\dag_0e^{-i\omega t + ikz}$, one can introduce the operators of annihilation and
creation of "photons in a medium" $\hat{c}_0$ and $\hat{c}^\dag_0$ \cite{fain} as $\hat{\textbf{E}}_0 = \textbf{e}_0 E_0 \hat{c}_0(\textbf{r},t), \hat{\bf{E}}^\dag_0 = \textbf{e}_0^* E_0
\hat{c}^\dag_0(\textbf{r},t)$. Here $\textbf{e}_0$ is a unit vector of the polarization of the field and  
$ \displaystyle
E_0 =  \sqrt{\frac{2\pi\hbar\omega^2\upsilon_{gr}}{c^2k}} $ is the normalization constant, 
where $\displaystyle  \upsilon_{gr} = \frac{2c^2k}{\partial(\omega^2n^2)/\partial\omega}$ is a group velocity. With this normalization of the field operators the
energy of the field in a volume V is given by $\hat{W} = \hbar \omega\left(V\hat{c}^\dag_0\hat{c}_0 + \frac{1}{2}\right)$ and their commutation relation
reads $\lbrack \hat{c}_0, \hat{c}^\dag_0\rbrack = \frac{1}{V}$. Therefore, these operators determine the number density of the photons in a certain state
$|\Psi_F\rangle$ of the field as $n_{Ph} = \langle \Psi_F |\hat{c}^\dag_0\hat{c}_0|\Psi_F\rangle$. This normalization is more convenient for field
propagation problems than the conventional normalization with $\lbrack \hat{c}_0, \hat{c}^\dag_0\rbrack = 1$ and $\hat{c}^\dag_0\hat{c}_0$ determining
the operator of the photon number.

A more realistic field consists of a certain number of modes propagating within a paraxial beam of a cross-sectional area $S_\perp$. If we keep the same notation
$\hat{c}_0$ for the field operators describing the field amplitude in the beam, their commutator becomes $\lbrack \hat{c}_0, \hat{c}^\dag_0\rbrack =
\Delta j / V$ where $\Delta j$ is the number of modes. The total photon flux density in the beam is then given by $Q = \upsilon_{gr} S_\perp \langle
\Psi_F |\hat{c}^\dag_0\hat{c}_0|\Psi_F\rangle$. 
It is convenient to go from a discrete set of modes to a continuous spectral interval $\Delta \omega \ll
\omega$. The density of states in a volume V is equal to $\eta = Vk^2/8\pi^3\upsilon_{gr}$ and the wave vectors of the modes constituting a beam
occupy the solid angle $\Delta o \approx 4\pi^2 / k^2 S_\perp$. As a result, we arrive at the following commutation relations for the operator of the
field amplitude and its spectral harmonics:
\begin{eqnarray}
&&\lbrack \hat{c}_0, \hat{c}^\dag_0 \rbrack = \displaystyle \Delta o\Delta\omega\eta = \frac{\Delta \omega}{2\pi S_\perp \upsilon_{gr}} \\
&& 
\lbrack \hat{c}_{0\omega^\prime}, \hat{c}^\dag_{0\omega^{\prime\prime}} \rbrack = \displaystyle \frac{1}{2\pi S_\perp\upsilon_{gr}}\delta(\omega^\prime -
\omega^{\prime\prime}) \\ &&
\lbrack \hat{c}_0(t^\prime), \hat{c}^\dag_0(t^{\prime\prime})\rbrack = \displaystyle \frac{\delta(t^\prime - t^{\prime\prime})}{S_\perp\upsilon_{gr}}
\end{eqnarray}
Here the spectral decomposition of the field amplitude operator is defined as
$
\hat{c}_0 = \int_{\Delta\omega}\hat{c}_{0\omega}e^{-i\omega t}d\omega
$.

The equation of motion for the field amplitude operator of each of the two signal fields has the same form as the wave equation for a classical field amplitude:
\begin{equation}
\left( \frac{\partial}{\partial t} + \upsilon_{gr}\frac{\partial}{\partial z}\right) \hat{c}_0 = \displaystyle
\frac{4\pi i \omega^2}{E_0 \partial(\omega^2n^2)/\partial\omega}\hat{\textbf{P}}_0\textbf{e}^*
\label{field1}
\end{equation}

Equation (\ref{field1}) includes all the relevant effects: linear dispersion determines the group velocity of the wave, whereas the slowly varying polarization amplitude $\hat{\textbf{P}}_0$ on the right-hand side includes nonlinearity, dissipation, and fluctuations.   At the boundary $z_b$ between the medium and the vacuum, the boundary condition for the field operator takes the form
\begin{equation}
\hat{c}_0(z_b)|_{vacuum} = \sqrt{\frac{\upsilon_{gr}}{c}}\hat{c}_0(z_b)|_{medium}
\label{bound}
\end{equation}
which satisfies the conservation of the Poynting flux. Eqs.~(\ref{rho}) are to be solved together with Eq.~(\ref{field1})
for both signal fields in order to determine the generated signal and noise.

In the four-wave mixing process depicted in Fig.~2, the total field consists of the four waves,
\begin{eqnarray}
\hat{\textbf{E}} &=& \textbf{e}_{(+)}E_{HF}e^{-i\omega_{HF}t + ik_{HF}z} + \textbf{e}_{(+)}E_{LF}e^{-i\omega_{LF}t + ik_{LF}z} \nonumber\\
&+& c.c.  + \textbf{e}_{(+)}E_0\hat{c}_{(+)}e^{-i\omega_{(+)}t + ik_{(+)}z} \nonumber\\
&+& \textbf{e}_{(-)}E_0\hat{c}_{(-)}e^{-i\omega_{(-)}t + ik_{(-)}z} + h.c. 
\end{eqnarray}
in which two strong classical pump fields at high and low frequencies (denoted as HF and LF) are resonant to the corresponding transitions between the Landau
levels, $\omega_{HF} = \omega_{41}$ and $\omega_{LF} = \omega_{21}$, whereas two signal fields are described by operators and their frequencies may
have a detuning, $\omega_{(+,-)} = \omega_{43, 32} \mp \Delta, \Delta \ll \omega_{+,-}$ satisfying the frequency-matching condition $\omega_{HF}
= \omega_{LF} + \omega_{(+)} + \omega_{(-)}$. We also assumed that $\omega_{(+)} \simeq \omega_{(-)} = \langle \omega \rangle$ in the normalization constant $E_0$. 

Solving the density-matrix equations (\ref{rho}) in the steady state and in linear approximation with respect to weak signal fields is reduced to a straightforward algebra. Optimal
conditions for the entanglement are realized when the Fermi level is close to the state 1 (n = -2) and the populations of all  states above
are low. This is possible when the magnetic field is strong enough, $k_B T \ll \hbar\omega_c$, and Rabi frequencies of the pump fields are
below saturation: $|\Omega_{HF,LF}| \ll \langle\gamma\rangle$. Here the Rabi frequencies are defined as $\displaystyle \Omega_{HF} = \frac{d^*_{14}E_{HF}}{\hbar}$, $\displaystyle
\Omega_{LF} = \frac{d^*_{12}E_{LF}}{\hbar}$, and we assume for simplicity that all scattering rates $\gamma_{mn}$ are of the same value
$\langle\gamma\rangle$. The latter assumption can be easily dropped once the relaxation rates are known for any particular sample. If, in addition,
the detuning is sufficiently large, $\langle\gamma\rangle \ll \Delta$, the populations of the excited states are mostly due to the Langevin
noise terms $\hat{F}_{(+,-)} \equiv \hat{F}_{32,43}$ in Eqs.~(\ref{rho}). Solving the density-matrix equations in the steady state and neglecting the terms of the order of
$(\langle\gamma\rangle/\Delta)^2$, we arrive at the following expression for the operator of the polarization amplitude at the frequency of the signal
fields:
\begin{equation}
\hat{\textbf{P}}_{(+,-)}  \approx \textbf{e}_{(+,-)}\left( \chi\hat{E}^\dag_{(-,+)}\mp
i\frac{d_{(+,-)}\hat{F}_{(+.-)}}{\Delta}\right)
\label{pol}
\end{equation}
where
\begin{equation}
\chi = \frac{Nd_{(+)}d_{(-)}}{\hbar\Delta}\frac{(\gamma_{21} + \gamma_{41})\Omega_{HF}\Omega^*_{LF}}{\gamma_{21}\gamma_{41}\gamma_{42}} \sim
\frac{Nd^2}{\hbar\Delta}\frac{\Omega^2_p}{\langle\gamma\rangle^2}
\end{equation}
and we denoted $\Omega^2_p = \Omega_{HF}\Omega^*_{LF}$ and $d = \hbar e \upsilon_F /\omega_{32}$.

Using the polarization (\ref{pol})) as a source in Eqs.~(\ref{field1}), we obtain the following coupled equations for the signal field operators:
\begin{equation}
\begin{array}{c}
\displaystyle
\left(\frac{\partial}{\partial z} + \frac{1}{\upsilon_{gr}}\frac{\partial}{\partial t}\right) \hat{c}_{(+)} = i\kappa \hat{c}^\dag_{(-)} +
\hat{G}_{(+)}\\
\displaystyle
\left(\frac{\partial}{\partial z} + \frac{1}{\upsilon_{gr}}\frac{\partial}{\partial t}\right) \hat{c}^\dag_{(-)} = -i\kappa^* \hat{c}_{(+)} +
\hat{G}^\dag_{(-)}\\
\end{array},
\label{field}
\end{equation}
where the coefficient of the parametric coupling is 
$$\displaystyle \kappa = 2\pi\chi\frac{\langle\omega\rangle^2}{c^2\langle k \rangle} 
$$ 
and the noise term \begin{equation}
 \displaystyle
\hat{G}_{(+,-)} = \mp 2\pi i \frac{\langle\omega\rangle^2}{c^2\langle k \rangle}\frac{d_{(+,-)}\hat{F}_{(+,-)}}{E_0\Delta}.
\label{G} 
\end{equation}
Here we again neglected a small
difference between the central frequencies of the signal fields in the pre-factors, assuming $\omega_{(+)} = \omega_{(-)} = \langle \omega \rangle$ and $\langle k \rangle = \langle \omega \rangle/c$.

 In the
optimal limit of $|\Omega_p| \ll \langle\gamma\rangle \ll |\Delta|$, the noise terms and the Raman scattering of the pump fields into the signal modes can be neglected and the solution for the fields exiting a layer of
thickness $L$ takes a particularly simple and transparent form:
\begin{eqnarray}
\hat{c}_{(+)}(L,t) &=& \displaystyle \cosh(\tau)\hat{c}_{(+)}\left(0,t-L/\upsilon_{gr}\right)\nonumber\\
&-&ie^{i\theta}\sinh(\tau)\hat{c}^\dag_{(-)}\left(0,t-L/\upsilon_{gr}\right) \nonumber\\
\hat{c}_{(-)}(L,t) &=& \displaystyle  \cosh(\tau)\hat{c}_{(-)}\left(0,t-L/\upsilon_{gr}\right) \nonumber\\
&-&ie^{i\theta}\sinh(\tau)\hat{c}^\dag_{(+)}\left(0,t-L/\upsilon_{gr}\right).  
\label{nonoise}
\end{eqnarray}
Here the parametric gain factor $\tau = |\kappa|L$ and $\kappa = |\kappa|e^{i\theta}$.

Eqs.~(\ref{nonoise}) clearly show the emergence of quantum correlations between the signal photons with opposite circular polarizations. In particular, consider the boundary condition at $z = 0$
corresponding to a completely uncorrelated state of vacuum fluctuations within the spectral bandwidth $\Delta\omega$. Then one can obtain from Eq.~(\ref{nonoise})
that the photon fluxes in two signal fields exiting the layer at $z = L$ are completely correlated:
\begin{equation}
\begin{array}{c}
\langle 0 | \hat{Q}_{(+)}(L) | 0\rangle = \displaystyle  \langle | \hat{Q}_{(-)}(L) | 0\rangle = \frac{\Delta\omega}{2\pi}\sinh^2\tau,  \\
\langle 0 | \left( \hat{Q}_{(+)}(L) - \hat{Q}_{(-)}(L)\right) ^2 | 0\rangle = 0 \\
\end{array}
 \label{flux}
\end{equation}
Here $\hat{Q}_{(+,-)}(L) = cS_\perp\hat{c}^\dag_{(+,-)}(L)\hat{c}_{(+,-)}(L)$ are operators of the photon fluxes. The correlated (+) and (-) photons can then be used to prepare the desired polarization-entangled states, e.g. the analog of the Bell states \cite{kwiat}. The second equation in (\ref{flux}) corresponds
to the Manley-Rowe relations for the parametric process. It also follows from Eq. (\ref{nonoise}) that the scheme could be used to amplify the light with a
nonclassical statistics or exchange the statistical properties between (+) and (-) photons. The magnitude of $\Delta \omega$  is likely to be limited by the bandwidth of a detection system.

If noise terms $\hat{G}_{+-}$ in Eq.~(\ref{field}) are taken into account, the field equations are still straightforward to solve, although the solution becomes more cumbersome. As a result, the photon fluxes in Eq.~(\ref{flux}) acquire additional noise terms:  
\begin{eqnarray} 
&&\langle0|\hat{Q}_{(+)}(L)|0\rangle \approx \frac{\Delta\omega}{2\pi}
\left(\sinh^2\tau + \right. \nonumber \\ &&  \left. \frac{\frac{\gamma_{(+)}}{\Delta}\Gamma_{(+)}(\sinh2\tau + 2\tau) + \frac{\gamma_{(-)}}{\Delta}\tilde{\Gamma}_{(-)}(\sinh2\tau - 2\tau)}{4|\kappa|}\right ) , \nonumber \\
&&\langle0|\hat{Q}_{(-)}(L)|0\rangle \approx \frac{\Delta\omega}{2\pi}  \left(\sinh^2\tau + \right. \nonumber \\
&& \left. \frac{\frac{\gamma_{(-)}}{\Delta}\Gamma_{(-)}(\sinh2\tau + 2\tau) + \frac{\gamma_{(+)}}{\Delta}\tilde{\Gamma}_{(+)}(\sinh2\tau - 2\tau)}{4|\kappa|}\right ) . \nonumber
\label{flux2}
\end{eqnarray}
Here the factors
\begin{eqnarray*}
\Gamma_{(+,-)} = 2\pi\frac{\langle\omega\rangle^2}{c^2\langle k \rangle}\frac{N_{4,3}|d_{(+,-)}|^2}{\hbar\Delta}, \\
\tilde{\Gamma}_{(+,-)} = 2\pi\frac{\langle\omega\rangle^2}{c^2\langle k \rangle}\frac{N_{3,2}|d_{(+,-)}|^2}{\hbar\Delta}
\end{eqnarray*}
are of the order of the parametric coupling terms:
\begin{equation}
\Gamma_{(+,-)}\sim\tilde{\Gamma}_{(+,-)}\sim\frac{\langle\omega\rangle^2}{c^2\langle k\rangle}\frac{Nd^2}{\hbar\Delta}\frac{\Omega^2_R}{\langle\gamma\rangle^2}\sim|\kappa|. \label{Gamma}
\end{equation}

From this solution one can
see that the noise contribution can be neglected if $|\Delta| \gg \langle \gamma \rangle$ provided the parametric gain is high enough: $\tau \ge 1$. For a weak amplification $\tau \ll 1$ the condition for a large signal to noise ratio is more stringent: $\Delta \gg \langle \gamma \rangle/\tau$. If this condition is not satisfied or if one of the states 2,3,or 4
acquires a large population, the noise is always comparable to or greater than the signal in the steady state. Then
the entangled photons can be generated only in the pulsed regime during the time of the order of a few relaxation times $1/\gamma$. This is usually the
case in resonant schemes of entangled photon generation in atomic vapors \cite{xiong,qamar}.

%%%%%%%%%%%%%%%%%%%%%%%%%%%%%%%%%%%%  figure 3 %%%%%%%%%%%%%%%%%%%%%%%%%%%%%
\begin{figure}[htb]
\centerline{
\includegraphics[width=7cm]{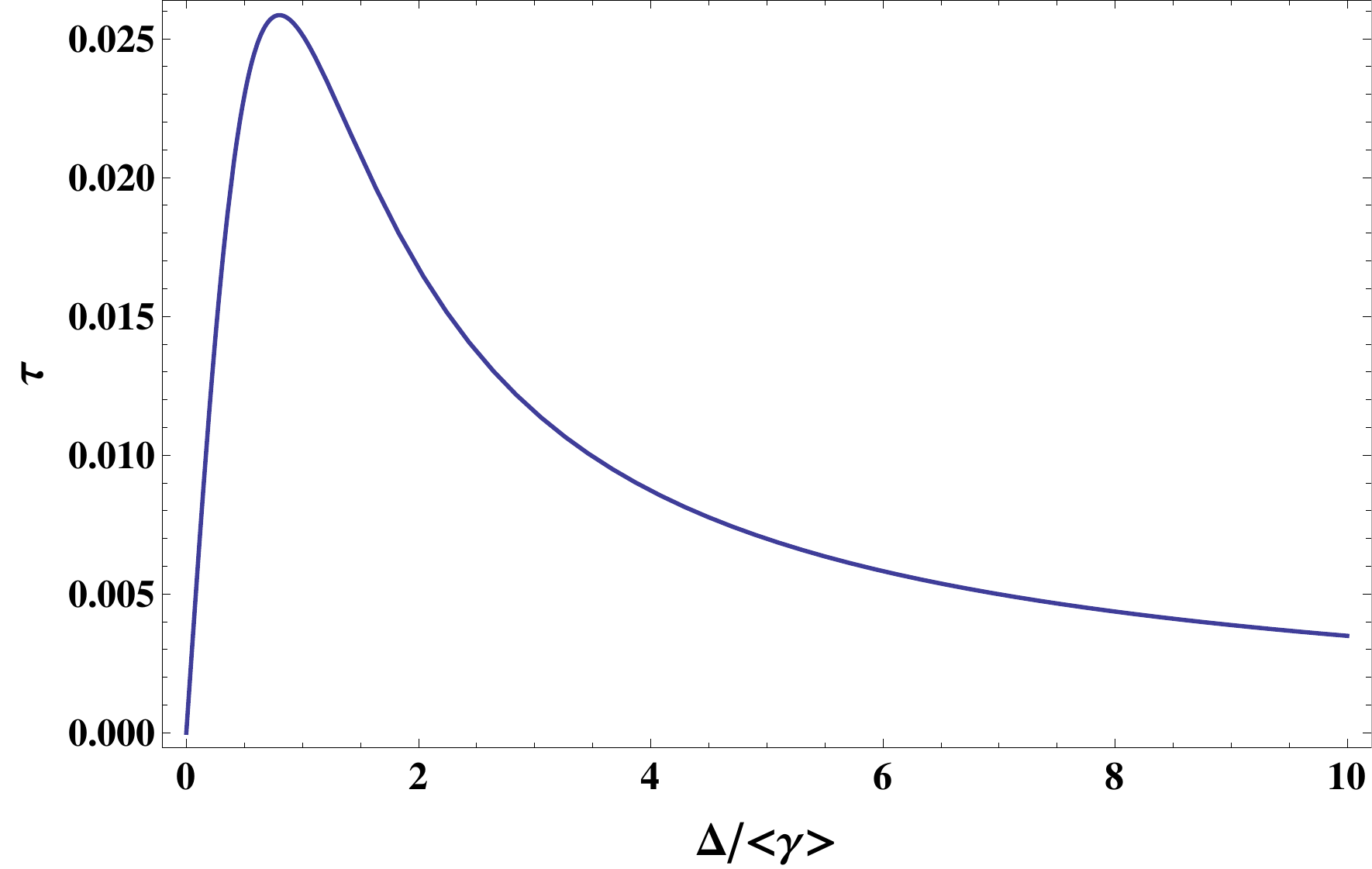}}
\caption{ Parametric gain $\tau$ per monolayer of graphene as a function of normalized detuning of the signal fields $\Delta/\langle \gamma \rangle$ for the pump field intensity $|\Omega_p|^2 = 0.1 \langle\gamma\rangle^2$. }
\end{figure}
%%%%%%%%%%%%%%%%%%%%%%%%%%%%%%%%%%%%%%%%%%%%%%%%%%%%%%%%%%%%%%%%%%%%%%%%%%%%

%%%%%%%%%%%%%%%%%%%%%%%%%%%%%%%%%%%%  figure 4 %%%%%%%%%%%%%%%%%%%%%%%%%%%%%
\begin{figure}[htb]
\centerline{
\includegraphics[width=7cm]{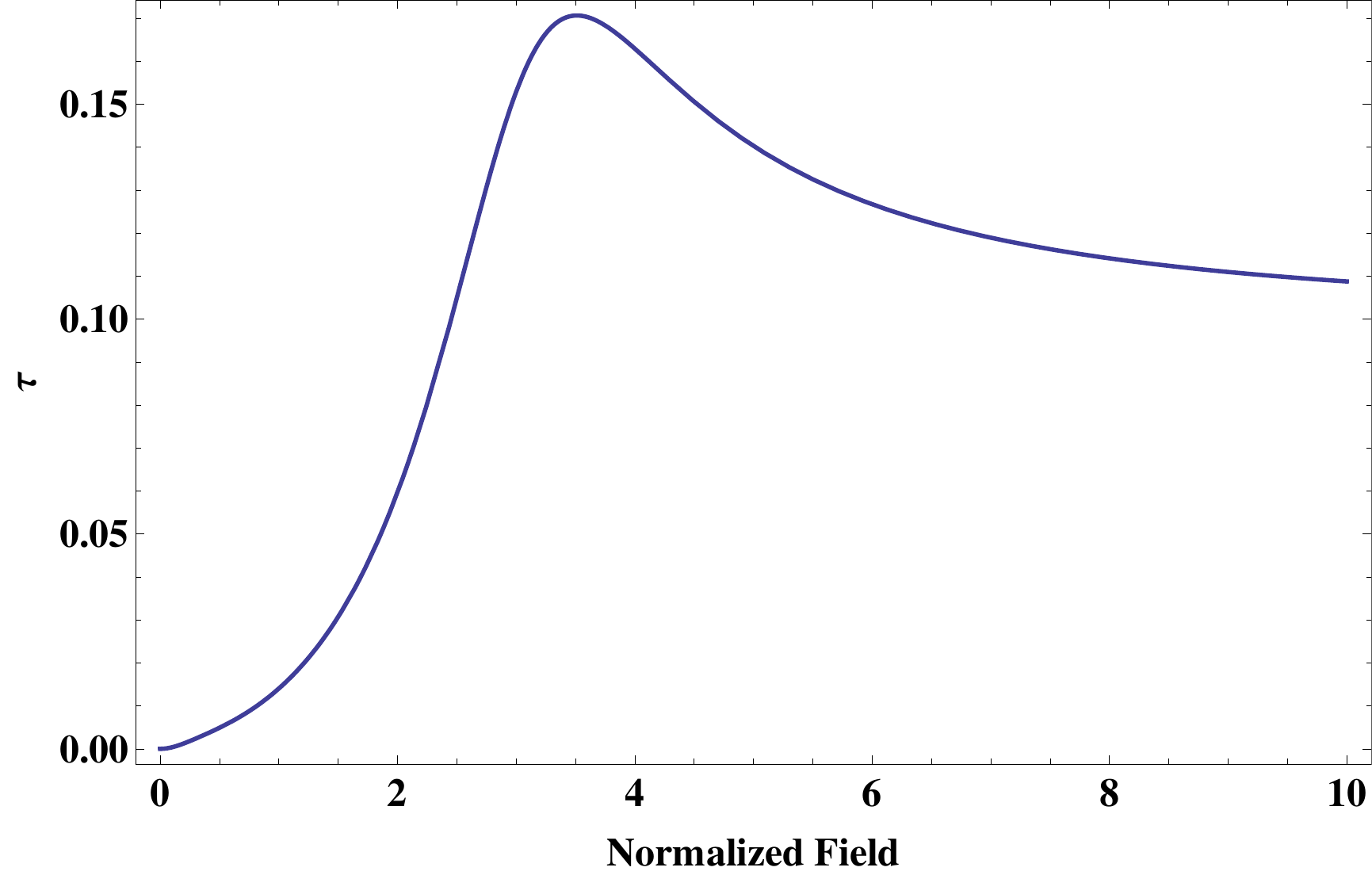}}
\caption{ Parametric gain $\tau$ per monolayer of graphene as a function of normalized strength of the pump field $|\Omega_p| / \langle\gamma\rangle$ for the detuning $\Delta = 10\langle \gamma \rangle$. }
\end{figure}
%%%%%%%%%%%%%%%%%%%%%%%%%%%%%%%%%%%%%%%%%%%%%%%%%%%%%%%%%%%%%%%%%%%%%%%%%%%%
The above analytic results were derived in the
 limit of $|\Omega_p| \ll \langle\gamma\rangle \ll \Delta$. In the general case the equations can be solved numerically, including the effects of the optical pumping of electrons to excited states and optical saturation. The resulting parametric gain $\tau$ per one monolayer of graphene is plotted in Figs. 3 and 4 as a function of the frequency detuning and the pump intensity. As seen from the figures, the magnitude of $\tau$ is around 0.01 for $\Delta \sim 10 \gamma \sim 100 \Omega_p$.  This corresponds to a photon flux of about $10^{-4} \Delta \omega/2\pi$. To increase the value of $\tau$ for a higher rate of the twin photon generation, one can use a stack of graphene monolayers or a thin layer of graphite. Recent studies have demonstrated that a graphite
layer consisting of $\sim 100$ monolayers maintains high carrier mobility and graphene-like Landau levels near the H-point of the Brillouin zone of graphite \cite{orlita2008,orlita2009}. The optimal
thickness is a tradeoff between the pump absorption and the desired output photon flux.

We showed that graphene placed in a magnetic field can serve as an  efficient, tunable source
of polarization-entangled photons in the mid-infrared and THz frequency range.  The proposed scheme can operate at high temperatures if the magnetic field is high enough to prevent thermal excitations,  $k_B T \ll \hbar\omega_c$.  A thin layer of graphene can be easily integrated with semiconductor laser
chips and optoelectronic circuits to make a compact setup. Similar mechanism of entangled photon generation could exist in topological insulators where the surface states have a massless dispersion and demonstrate a similar pattern of Landau levels $\propto \sqrt{Bn}$ and magneto-absorption  \cite{topins1}. Interestingly, the band velocity $\upsilon_F \sim 8.5\times 10^7$ cm/s for surface states in Bi$_{0.91}$Sb$_{0.09}$ inferred from measurements in \cite{topins1} is very close to the one in graphene, which suggests an optical nonlinearity of similar strength. We hope that our results will stimulate active experimental research in nonlinear and quantum optics of graphene and other materials with similar electronic properties. 

The authors are grateful to Valery Vdovin and Maria Erukhimova for helpful discussions. This work has been supported in part by NSF Grants OISE-0968405 and EEC-0540832, by the NHARP Project No. 003658-0010-2009, and by the Federal Target Program ''Research and Development in
Priority Directions of Development of Russia Scientific-Technological Complex'' 
(grant No. 07.514.11.4162).


\begin{thebibliography}{20}
\bibitem{kwiat} P. G. Kwiat, K. Mattle, H. Weinfurter, A. Zeilinger, A. V. Sergienko and Y. Shih, Phys. Rev. Lett. 75, 4337 (1995).
\bibitem{nature2012} J. Yin, J. Ren, H. Lu, Y. Cao, H. Yong, et al., Nature 488, 185 (2012).
\bibitem{fan-opn07} J. Fan, A. Migdall and L. Wang, Opt. Photon. News, March 2007, 26-33.
\bibitem{stevenson} R. M. Stevenson, R. J. Young, P. Atkinson, K. Cooper, D. A. Ritchie and A. J. Shields, Nature 439, 179Ð182 (2006).
\bibitem{mohan} A. Mohan, M. Felici, P. Gallo, B. Dwir, A. Rudra, et al., Nature Phot. 4, 302 (2010).
\bibitem{dousse} A. Dousse, J. Suffczynski, A. Beveratos, O. Krebs, A. Lemaitre, et al., Nature 466, 217 (2010).
\bibitem{castroneto} A. H. Castro Neto, F. Guinea, N. M. R. Peres, K. S. Novoselov, and A. K. Geim,
Rev. Mod. Phys. 81, 109162 (2009).
\bibitem{nair} R. R. Nair, et al., Science 320, 1308 (2008).
\bibitem{sadowski} M. L. Sadowski, G. Martinez, M. Potemski, C. Berger and W. A. de Heer,
 Phys. Rev. Lett. 97, 266405 (2006).
\bibitem{abergel2007} D.S.L. Abergel and V. I. Fal'ko,  Phys. Rev. B 75, 155430 (2007).
\bibitem{prl} X. Yao and A. Belyanin,  Phys. Rev. Lett. 108, 255503 (2012).
\bibitem{kono2012} L. G. Booshehri, et al.  Phys. Rev. B 85, 205407 (2012).
\bibitem{orlita2008} M. Orlita, et al., Phys. Rev. Lett. 101, 267601 (2008).
\bibitem{orlita2009} M. Orlita, et al.  Phys. Rev. Lett. 102, 166401 (2012).
\bibitem{crassee} I. Crassee, et al.  Nature Phys. 7, 48 (2011).
\bibitem{landau} L. D. Landau and E. M. Lifshitz, {\it Statistical Physics, Part 1}. Butterworth-Heinemann, Oxford, 1980.
\bibitem{rytov} S. M. Rytov, Yu. A.  Kravtsov,  and V. I. Tatarskii, {\it Principles of Statistical Radiophysics 3: Elements of random fields}, Springer-Verlag, Berlin 1989. 
\bibitem{fain} V. M. Fain and Ya. I. Khanin, {\it Quantum electronics, Vol. 1.} The MIT Press, Cambridge, MA 1969.
\bibitem{xiong} H. Xiong, M. O. Scully and M. S. Zubairy, Phys. Rev. Lett. 94, 023601 (2005).
\bibitem{qamar} S. Qamar, M. Al-Amri and M. S. Zubairy, Phys. Rev. A 79, 013831 (2009).
\bibitem{topins1} A. A. Schafgans, K. W. Post, A. A., Taskin, Y. Ando, X. L. Qi, B. C. Chapler, and D. N. Basov,  Phys. Rev. B. 85, 195440 (2012). 



\end{thebibliography}
\end{document}